\documentclass[journal, 11pt, onecolumn]{IEEEtran}

\usepackage[noadjust]{cite}

\usepackage[pdftex]{graphicx}               
\graphicspath{{../pdf/}{../jpeg/}}
\DeclareGraphicsExtensions{.pdf,.jpeg,.png}
\usepackage{amsmath}                        % Enhanced mathematical typesetting
\usepackage{amssymb}                        % Additional mathematical symbols
\usepackage{mathtools}                      % Additional math environments and tools
\usepackage{algorithm}                      % Algorithm typesetting
\usepackage{algpseudocode}                  % Pseudocode typesetting
\usepackage{multirow}                       % Multi-row cells in tables
\usepackage{booktabs}                       % Enhanced table formatting
\usepackage{siunitx}                        % Typesetting of units
\usepackage{url}                            % URL typesetting

\usepackage{tikz}                           % Package for creating graphics programmatically
\usepackage{pgfplots}                       % High-quality plots and diagrams
\usepgfplotslibrary{colorbrewer}
\usetikzlibrary{intersections}
\usepgfplotslibrary{polar}
\usetikzlibrary{shapes, arrows.meta}
\pgfplotsset{compat=newest}

\definecolor{color1}{RGB}{0, 0, 127}         % Dark blue
\definecolor{color2}{RGB}{0, 63, 255}        % Light blue
\definecolor{color3}{RGB}{63, 255, 191}      % Cyan
\definecolor{color4}{RGB}{255, 191, 63}      % Yellow
\definecolor{color5}{RGB}{255, 63, 0}        % Orangec
\definecolor{color6}{RGB}{191, 0, 0}         % Dark red
\definecolor{darkgray}{RGB}{50, 50, 50}      % Dark gray

\usepackage{array}
\newcolumntype{L}[1]{>{\raggedright\let\newline\\\arraybackslash\hspace{0pt}}m{#1}}
\newcolumntype{C}[1]{>{\centering\let\newline\\\arraybackslash\hspace{0pt}}m{#1}}
\newcolumntype{R}[1]{>{\raggedleft\let\newline\\\arraybackslash\hspace{0pt}}m{#1}}

\usepackage{microtype}  % This package improves justification and reduces hyphenation
\newcommand{\diag}{\mathop{\bf diag}}

\newcommand{\argmin}{\mathop{\rm argmin}}

\newcommand{\eg}{{\it e.g.}}
\newcommand{\ie}{{\it i.e.}}

\newcommand{\R}{\mathbb{R}}

\newcommand{\vct}[1]{\boldsymbol{#1}}
\newcommand{\mtx}[1]{\boldsymbol{#1}}
\newcommand{\vg}{\vct{g}}

\newcommand{\vk}{\vct{k}}

\newcommand{\vn}{\vct{n}}

\newcommand{\vr}{\vct{r}}

\newcommand{\vu}{\vct{u}}

\newcommand{\vw}{\vct{w}}
\newcommand{\vx}{\vct{x}}
\newcommand{\vy}{\vct{y}}

\newcommand{\valpha}{\vct{\alpha}}

\newcommand{\vphi}{\vct{\phi}}

\newcommand{\vtheta}{\vct{\theta}}

\newcommand{\mU}{\mtx{U}}

\newcommand{\mX}{\mtx{X}}
\newcommand{\mY}{\mtx{Y}}

\newcommand{\mPsi}{\mtx{\Psi}}

\begin{document}

\title{\rule{\textwidth}{2pt} \textbf{Single-shot Tomography of Discrete Dynamic Objects}\\[-1ex] \rule{\textwidth}{2pt} }

\author{Ajinkya~Kadu$^{1,2}$, 
        Felix~Lucka$^{2}$,
        and~Kees~Joost~Batenburg$^{2,3}$ \\[2ex]
        {\small $^{1}$Electron Microscopy for Material Sciences, University of Antwerp, 2020 Antwerp, Belgium \\[0.2ex]
        $^{2}$Computational Imaging, Centrum Wiskunde and Informatica, 1098 XG Amsterdam, Netherlands \\[0.2ex]
        $^{3}$Leiden Institute of Advanced Computer Science, 2333 CA Leiden, Netherlands \par}
}

% \markboth{Journal of \LaTeX\ Class Files,~Vol.~14, No.~8, August~2015}%
% {Shell \MakeLowercase{\textit{et al.}}: Bare Demo of IEEEtran.cls for IEEE Journals}

\maketitle

%%%%%%%%%%%%%%%%%%%%%%%%%%%%%%%%%%%%%%%%%%%%%%%%%%%%%%%%%%%%%%%%%
%% ABSTRACT
%%%%%%%%%%%%%%%%%%%%%%%%%%%%%%%%%%%%%%%%%%%%%%%%%%%%%%%%%%%%%%%%%
\begin{abstract}
This paper presents a novel method for the reconstruction of high-resolution temporal images in dynamic tomographic imaging, particularly for discrete objects with smooth boundaries that vary over time. Addressing the challenge of limited measurements per time point, we propose a technique that synergistically incorporates spatial and temporal information of the dynamic objects. This is achieved through the application of the level-set method for image segmentation and the representation of motion via a sinusoidal basis. The result is a computationally efficient and easily optimizable variational framework that enables the reconstruction of high-quality 2D or 3D image sequences with a single projection per frame. Compared to current methods, our proposed approach demonstrates superior performance on both synthetic and pseudo-dynamic real X-ray tomography datasets. The implications of this research extend to improved visualization and analysis of dynamic processes in tomographic imaging, finding potential applications in diverse scientific and industrial domains.
\end{abstract}


% sections
% \input{sections/introduction}
%%%%%%%%%%%%%%%%%%%%%%%%%%%%%%%%%%%%%%%%%%%%%%%%%%%%%%%%%%%%%%%%%
%% Introduction
%%%%%%%%%%%%%%%%%%%%%%%%%%%%%%%%%%%%%%%%%%%%%%%%%%%%%%%%%%%%%%%%%
\section{Introduction}
\label{sec:Intro}

Computational imaging (CI) technologies, such as X-ray computed tomography (CT), magnetic resonance imaging (MRI) and ultrasound (US), facilitate the generation of detailed, high-resolution images of static object interiors. This ability stems from the moderately ill-posed nature of the underlying inverse problems that seek to determine an object's internal structure based on external characteristic measurements\cite{natterer2001mathematics}. Consequently, the collected and processed data yield a relatively high confidence level, providing critical insights into the structures of diverse objects and systems. In biomedical applications, imaging techniques have been widely used for anatomy, disease diagnosis, and the development of novel treatments and therapies\cite{li2014deep, dai2020ct, furukawa2004cross}. In the broader field of science, CI has shown its relevance in areas such as material sciences, where it aids in understanding the properties and structures of materials\cite{midgley2009electron}; geosciences, for the analysis of geological formations and subsurface structures\cite{bai2010crustal}; and astrophysics, where it contributes to the study of celestial bodies and phenomena\cite{akiyama2019first}. These technologies have instigated a paradigm shift in imaging and continue to enhance our understanding of the world across various disciplines.

Moreover, CI technologies extend beyond static object imaging, with an increasing focus on capturing dynamic processes at various scales to gain valuable insights into underlying mechanisms and a more profound understanding of complex systems. In medical imaging, CT and MRI have been employed to observe dynamic blood flow behavior, visualize medical intervention impacts, and track disease progression, such as in cancer cases\cite{feher2017quantitative, craig2020tumour, zhang2017integrating}. Furthermore, these technologies can be applied to material science, enabling the study of material dynamic behavior\cite{skorikov2019quantitative, ruhlandt2017four, calta2018instrument, wen2019time}.

%-----------------------------------------------------------------
\begin{figure}[t]
    \centering
    \includegraphics[width=0.5\columnwidth]{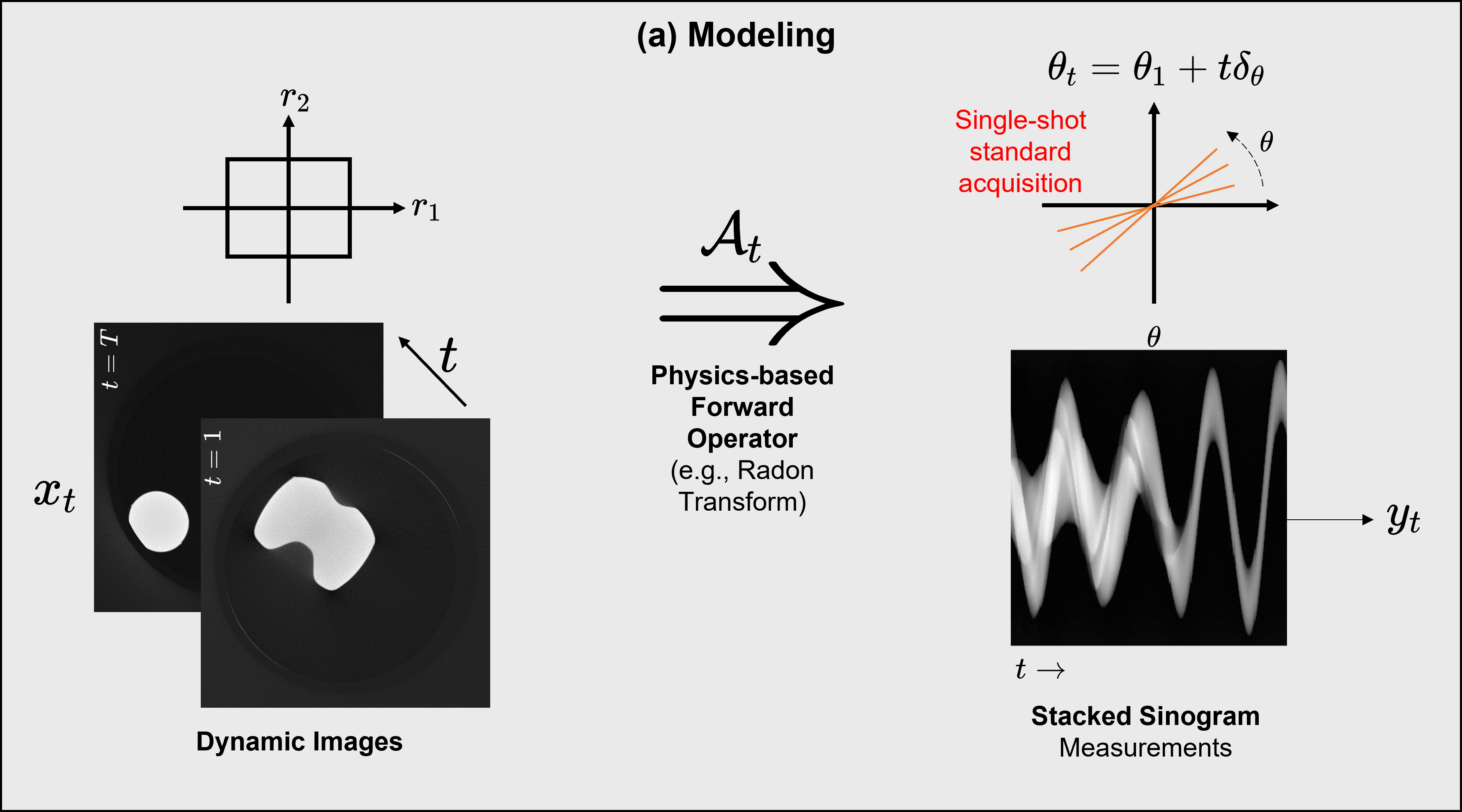} \\[0.5ex]
    \includegraphics[width=0.5\columnwidth]{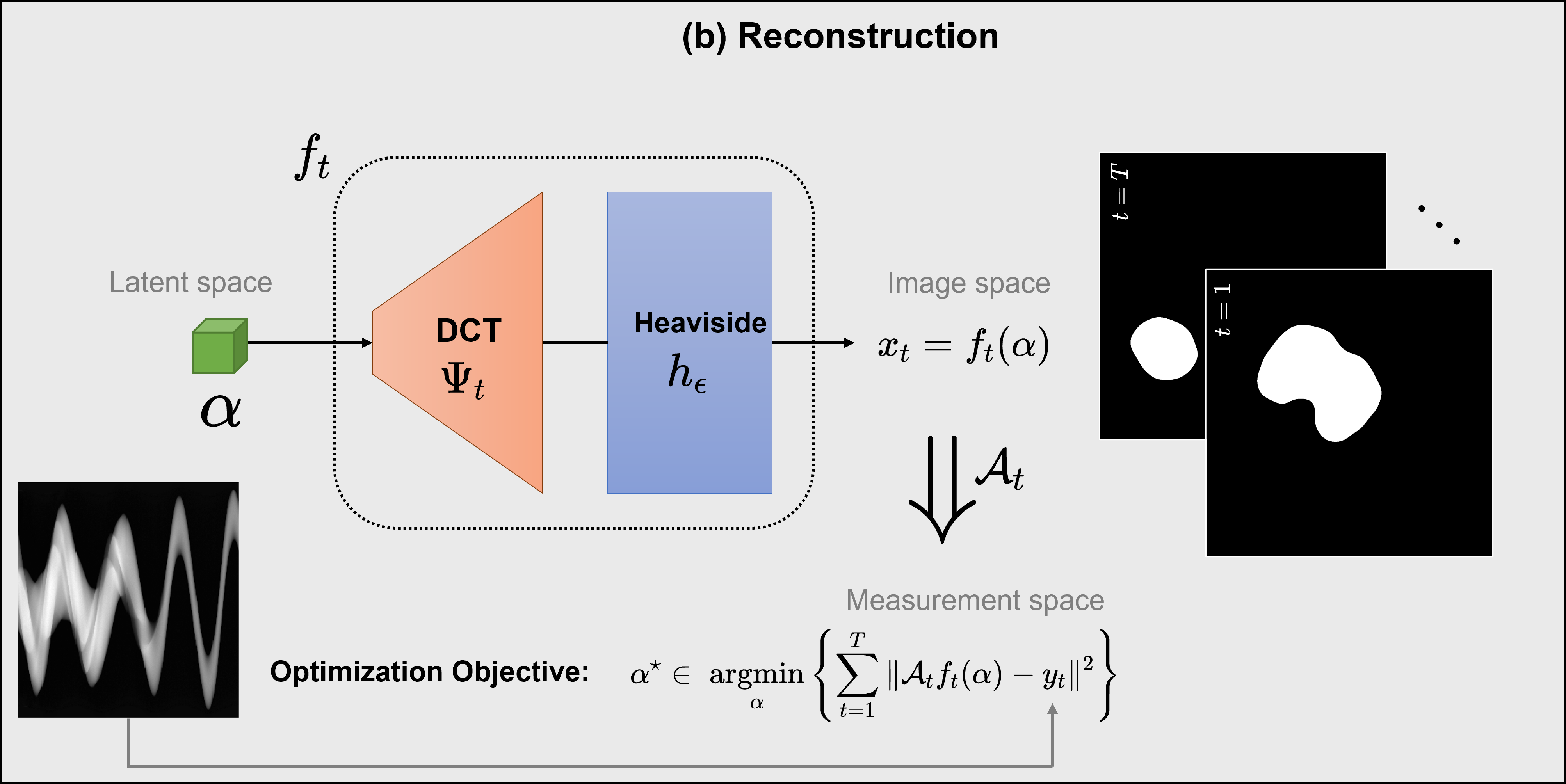}
    \caption{Sketch of the proposed algorithm: (a) Visualizing the interplay between dynamic images and measurements through the forward operator $\mathcal{A}_t$, where a tomographic scenario is employed, and the Radon transform represents the forward operator. (b) The reconstruction phase is characterized by addressing an optimization challenge that aims to minimize the least-squares discrepancy within the measurement domain, while concurrently associating the image with the DCT coefficients $\boldsymbol{\alpha}$ via the transformative function $f_t$.}
    \label{fig:teaser}
    \vspace{-1em}
\end{figure}
%-----------------------------------------------------------------
\begin{figure*}[!t]
    \centering
    \includegraphics[width=0.8\textwidth]{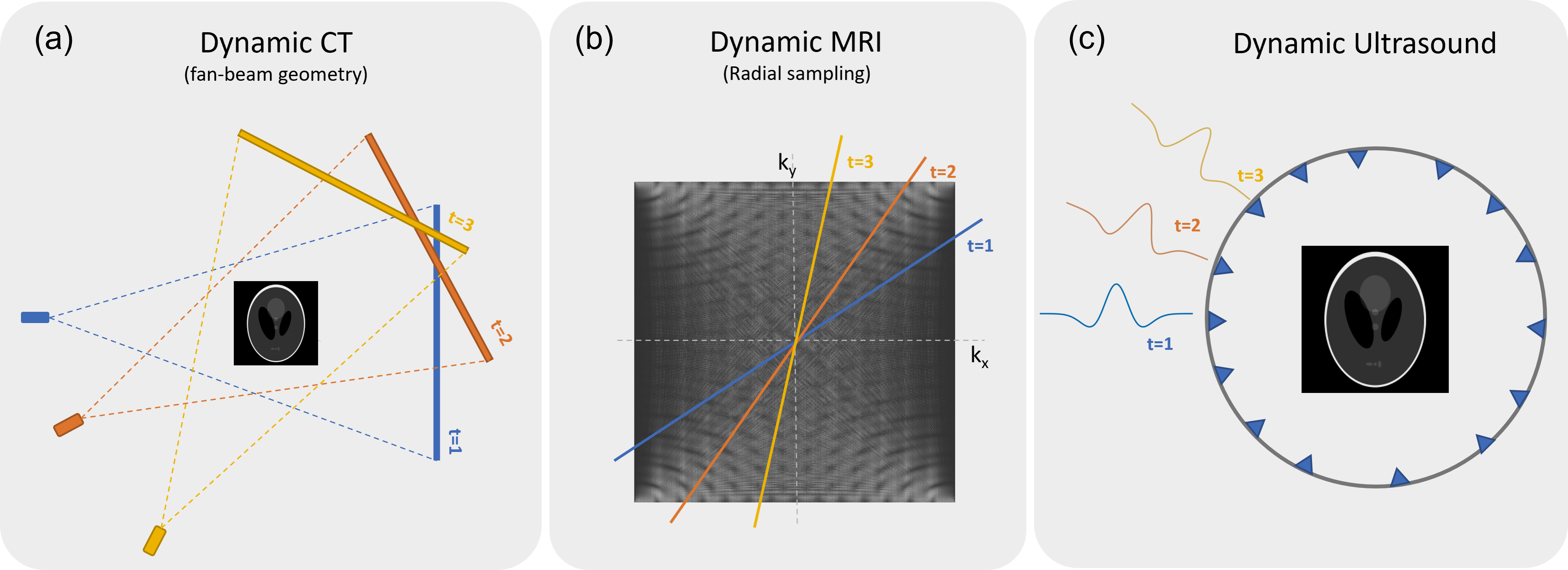}
    \caption{Single-shot dynamic imaging scenarios for various imaging modalities. (a) Dynamic CT: Fan-beam geometry with sampling from three time slices, each represented by a single angle, (b) Dynamic MRI: Radial sampling in k-space for a Shepp-Logan phantom, illustrating three radial lines corresponding to different time slices, (c) Dynamic US: A circular array of detectors and sources surrounding the phantom, capturing three time samples with a single detector emitting an acoustic signal at each time instance.}
    \label{fig:single_shot_exp}
\end{figure*}
%-----------------------------------------------------------------

Imaging dynamic processes with CI modalities presents challenges often arising from the sequential acquisition of data set dimensions over time. For instance, in CT, the angular dimension is acquired by rotating either the object or the X-ray source-detector pair\cite{kalender2006x}. In MRI, k-space lines are acquired sequentially to collect the complete data\cite{hutchinson1988fast}. In US, acoustic waves are emitted sequentially from various source locations to acquire data\cite{jensen2005ultrasound}. Standard image reconstruction algorithms can produce artifacts in the final image if they fail to account for motion-induced data inconsistencies\cite{zaitsev2015motion, wang2008outlook, nelson2000sources}.

The occurence of such artifacts due to the dynamic evolution of the object depends strongly on the rate of change of the object with respect to a fully 3D image acquisition process. For dynamic processes where changes are relatively small compared to the full acquisition time, a \emph{snapshot} strategy is typically employed, capturing a single state of the object for each full acquisition procedure. In synchrotron tomography, for example, the sample can be rotated over 20 times per second, offering a temporal resolution above 20Hz in the snapshot-based reconstruction \cite{maire2016, buurlage2019}. For scenarios where the object moves in a periodic manner, such as the human heart and lungs, \emph{gating} procedures can be used that combine measurements from different cycles to jointly form a complete measurement for each separate time point \cite{song2007}.

In cases where the object exhibits strong, non-periodic dynamics during a single full acquisition cycle, the snapshot and gating strategies cannot be effectively applied. One can choose to reduce the number of acquisitions to improve temporal resolution, but this can result in missing data artifacts. To address this, researchers have devised innovative solutions that supplement the missing information using image models describing spatial image characteristics. One prevalent method employs sparse image models, representing images as compositions of simple, sparsely occurring building blocks\cite{van2015compressed, leary2013compressed}. Another approach, which has gained traction in recent years, utilizes deep learning-based models, where machine learning algorithms are trained on extensive datasets to discern patterns and relationships between image data and content\cite{ ardila2019end, baguer2020computed, adler2018learned}. Additionally, discrete tomography serves as an alternative method, providing a mathematical framework for image reconstruction that models the image as a set of discrete, quantized values\cite{batenburg2011dart, zhuge2015tvr, dorn2006level, kadu2017parametric}.

In certain scenarios, the object under investigation can only be assumed static during a single measurement (refer to Figure \ref{fig:single_shot_exp} for data acquisition). This represents the extreme case of inverse problem ill-posedness but provides optimal temporal resolution. Here, we refer to this case as \textit{single-shot imaging}. Conventional image models are typically insufficient for supplementing the missing information in such situations. Consequently, researchers have developed a variety of spatio-temporal image models that incorporate both spatial and temporal image aspects (see review paper \cite{hauptmann2021image}). These models can be highly complex, ranging from mathematical models incorporating dynamic process information to machine learning models learning the relationships between image data and content. For example, Niemi et al. \cite{niemi2015dynamic} proposed a dynamic X-ray tomography method using a spacetime level set, which sought to simultaneously reconstruct the shape and motion of an object. This work demonstrated a promising approach in handling dynamic processes in X-ray tomography but was largely applicable to cases where more than 1 tomographic projections are available at every snapshot.

This paper focuses on the specific case of discrete objects with smooth boundaries that vary smoothly over time, which holds relevance for numerous applications in materials science \cite{batenburg20093d, zhuo2022} and engineering \cite{browne1998, zeilinga2021}. The paper's contribution is a novel approach called \textit{Dynamic Shape Sensing} for single-shot tomographic imaging of such objects, combining the level-set method and compressed motion sensing to reconstruct the spatiotemporal motion of discrete objects (the workflow described in Figure~\ref{fig:teaser}). Despite the non-convex nature of the loss function, a gradient descent method is employed to obtain the numerical solution. The effectiveness of the approach is demonstrated through simulated and experimental data obtained from X-ray tomography datasets. This innovative method offers potential for advancing the field of computational imaging and expanding its applicability across various domains, furthering our understanding of complex systems and processes that require high temporal resolution and accurate reconstructions.

% \input{sections/methods}
%%%%%%%%%%%%%%%%%%%%%%%%%%%%%%%%%%%%%%%%%%%%%%%%%%%%%%%%%%%%%%%%%
%% Methods
%%%%%%%%%%%%%%%%%%%%%%%%%%%%%%%%%%%%%%%%%%%%%%%%%%%%%%%%%%%%%%%%%
\section{Methods}
\label{sec:Methods}

In this section we first introduce the dynamic tomographic inverse problem, followed by the common approach for modeling regularized dynamic reconstruction. This approach is highly computationally challenging, imposing strong limitations on its applicability. We then move on to our main contribution, introducing \emph{dynamic shape sensing}, a generalization of \emph{compressed shape sensing}, which is less limited in those regards. 

%%%%%%%%%%%%%%%%%%%%%%%%%%%%%%%%%%%%%%%%%%%%%%%%%%%%%%%%%%%%%%%%%
%% Dynamic Tomographic Inverse Problem
%%%%%%%%%%%%%%%%%%%%%%%%%%%%%%%%%%%%%%%%%%%%%%%%%%%%%%%%%%%%%%%%%
\subsection{Dynamic Tomographic Inverse Problem}
In dynamic tomographic imaging, accurately modeling the relationship between the image and the measurements is crucial. The forward model defines this relationship by describing the physical process that transforms the image into the measurement data. In this study, the forward model is expressed as 
\begin{equation}
    \vy_t = \mathcal{A}_t (\vx_t) + \vct{\varepsilon}_t , \label{eq:FwdModel}
\end{equation}
where $\mathcal{A} = \{ \mathcal{A}_1, \dots, \mathcal{A}_T \}$, with each $\mathcal{A}_t : \mathbb{X} \rightarrow \mathbb{Y}$  representing a time-series of linear or non-linear \textit{forward operators} mapping from the image space~$\mathbb{X}$ to the measurement space~$\mathbb{Y}$. $\mX = \{ \vx_1 , \dots , \vx_T \}$  denotes the time series of images to be recovered from the time-series of measurements $\mY = \{\vy_1, \dots, \vy_T \}$. We assume that each measurement is corrupted by additive white Gaussian noise (AWGN) $\vct{\varepsilon}_t$. One solution of \eqref{eq:FwdModel} for each $t$ separately can be formulated by considering the constraint least-squares problem
\begin{align}
    \vx^\star_t \in \argmin_{\vx_t \in \mathcal{C}} \, \Big\lbrace \| \mathcal{A}_t (\vx_t) - \vy_t \|^2 \Big\rbrace , \label{eq:LeastSquares}
\end{align}
where $\| \cdot \|_2$ is the Euclidean norm and $\mathcal{C}$ denotes a-priori constraints on $\vx$, such as box-constraints on the image intensities $0 \leq \vx(r) \leq 1$.

In X-ray CT, the forward operator models the projection of an object's X-ray absorption coefficients along a set of angles using a mathematical operation known as the Radon transform \cite{natterer2001mathematics}. This operator maps the temporal images, which represent the X-ray absorption coefficients of the object over time, to the measurements, which represent the decay in the X-ray intensities at the detector. Specifically, the Radon transform is a mathematical integral defined over the image domain $\Omega$ and takes the form:
\begin{align}
    \mathcal{T}[\vx](s,\vtheta) = \int_{\Omega} \vx(\vr) \delta(s - \langle \vr, \vn(\vtheta) \rangle ) \, \mathrm{d} \vr,
\end{align}
where $s$ and $\vtheta$ are the distance and Euler angles of the projection, respectively. Here, $\vr$ represents the spatial coordinates of the image, $\delta$ is the Dirac-delta function, and $\vn(\theta)$ is the unit vector along the direction $\vtheta$. 
In the dynamic case, with $d = 2$ and parallel-beam geometry, $\vtheta$ is sampled in time, leading to
\begin{align*}
    \mathcal{A}_t(\vx_t) =  \mathcal{T}[\vx_t](s,\vtheta_t = \vtheta_{1} + t \, \delta_{\vtheta})
\end{align*}
where $\vtheta_1$ is an angular direction at an initial time point $t = 1$, and $\delta_{\vtheta}$ represents the difference between two consecutive angular directions. Hence, for each time point $t$, a 2D image $\vx_t$ is mapped to a 1D function $\vy_t$, which leads to a severely ill-posed inverse problem. Dynamic forward models for MRI and US have been provided in Appendix~\ref{sec:ForwardModels} for comparison. These models are essential for accurate and efficient image reconstruction, as they provide a theoretical foundation for the CI pipeline.

%%%%%%%%%%%%%%%%%%%%%%%%%%%%%%%%%%%%%%%%%%%%%%%%%%%%%%%%%%%%%%%%%
%% Regularized Dynamic Reconstruction
%%%%%%%%%%%%%%%%%%%%%%%%%%%%%%%%%%%%%%%%%%%%%%%%%%%%%%%%%%%%%%%%%
\subsection{Regularized Dynamic Reconstruction}
\label{sec:Methods:Regularized}
The simplest approach for dynamic reconstruction is to assume that the object does not change much within time intervals of length $B$ called \textit{bins} and reconstruct it from all the data acquired during a bin as if it were static. Mathematically, binning translates into imposing the equality constraints 
\begin{equation}
    \begin{gathered}
        \vx_1 = \vx_2 = \cdots = \vx_B; \\
        \vx_{B+1} = \vx_{B+2} = \cdots = \vx_{2 B}; \\
        \vdots \\
        \vx_{T-B+1} = \vx_{T-B+2} = \cdots = \vx_{T}.
    \end{gathered}
    \label{eq:binning}
\end{equation}
and solving the optimization problem
\begin{equation}
    \begin{aligned}
     \vx_{[i]}^\star & \in \argmin_{\vx \in \mathcal{C}} \Bigg\lbrace \sum_{t= (i-1)B +1}^{iB} \| \mathcal{A}_t ( \vx ) - \vy_t \|^2 + \mathcal{R}(\vx) \Bigg\rbrace,
    \end{aligned}
    \label{eq:Static}
\end{equation}
where $i$ denotes the bin. The regularization functional $\mathcal{R}(\vx)$ imposes additional prior information about the spatial structure of the object. In this work, we want to model piece-wise homogeneous objects with smooth boundaries and will use the Total-Variation (TV) functional $\mathcal{R}(\vx) = \| \nabla \vx \|_1$. Although this method enables obtaining the solution bin-by-bin, it is not flexible with respect to to temporal variations in $\vx_t$ that happen within one bin and such violations of the equality constraints introduce motion artifacts into the reconstruction.

To overcome this limitation, a more sophisticated approach relaxes the equality constraints by penalizing differences between consecutive frames through a suitable norm, such as the $\ell_2$ norm. While this promotes smooth motions, the reconstructions for different $t$ are now coupled, which means that we need to solve for the entire series of $\mX$ simultaneously:
\begin{equation}
    \begin{split}
        \mX^* &\in \argmin_{\mX \in \mathcal{C}} \Bigg\lbrace \sum_{t = 1}^{T} \| \mathcal{A}_t \! \left( \vx_t \right) - \vy_t \|^2 \, + \, \alpha \, \| \nabla \vx \|_1 \Bigg. \\
        & \qquad \qquad \Bigg. + \, \beta \sum_{t=1}^{T-1} \| \vx_{t+1} - \vx_{t} \|^2 \Bigg\rbrace .
    \end{split}
    \label{eq:TemporalRegularization}
\end{equation}

The integration of a motion model $M$ into the reconstruction can further improve the accuracy and reliability of the results. Instead of assuming $\vx_{t} \approx \vx_{t+1}$ as in \eqref{eq:TemporalRegularization}, we use $\vx_{t} \approx M(\vx_{t+1}, \vu_t)$, where $\vu_t$ are variables describing motion (note that in this formulation, $M$ describes motion backwards in time). Incorporating this information into our optimization problem leads to
\begin{equation}
    \begin{split}
        \mX^* &\in \argmin_{\mX \in \mathcal{C}} \bigg\lbrace \sum_{t = 1}^{T}  \| \mathcal{A}_t \! \left( \vx_t \right) - \vy_t \|^2 \, + \, \alpha \| \nabla \vx \|_1 \bigg. \\
        & \bigg. \qquad \qquad + \, \beta  \sum_{t=1}^{T-1} \| M(\vx_{t+1}, \vu_t) - \vx_t \|^2 \bigg\rbrace
    \end{split}
    \label{eq:MotionRegularization}
\end{equation}
Examples of such motion models include the optical flow model, $M(\vx_{t+1}, \vu_t) = \vx_{t+1}(\vr + \vu_t)$, or its linearization for small displacements, $\vu_t$, given as $M(\vx_{t+1}, \vu_t) \approx \vx_{t+1} + (\nabla \vx_{t}) \vu_t$.

However, if the motion parameters, $\vu$, are unknown, we must estimate them from the data, too. This leads to a joint image reconstruction and motion estimation problem, which is typically non-convex and requires appropriate regularization on $\vu$ to ensure stability and accuracy. The particular formulation we will consider here will be called TV-TV-OF:
\begin{equation}
    \begin{aligned}
        \mX^\star, \mU^\star &= \argmin_{\mX \in \mathcal{C}, \mU} \bigg\lbrace \sum_{t = 1}^{T} \| \mathcal{A}_t (\vx_t) - \vy_t \|^2  + \alpha \, \| \nabla \vx_t \|_1 \bigg. \\
    & \bigg. \quad + \beta \, \| \vx_{t+1}  + \left( \nabla \vx_t \right) \vu_t - \vx_{t} \|^2 + \gamma \, \| \nabla \vu_t \|_1  \bigg\rbrace
    \end{aligned}
    \label{eq:TVOptFlow}
\end{equation}
Using TV regularization on the displacement fields $\vu_t$ is a common model in optical flow estimation\cite{BuDiFr15} and tries to decompose the domain into areas in which the displacement field is constant. More information on the TV-TV-OF model can be found in \cite{Di15,BuDiSc16,burger2017variational,lucka2018enhancing}. A major drawback of TV-TV-OF is its computational complexity and substantial memory requirements, which renders its implementation for high-resolution 3D scenarios extremely challenging \cite{lucka2018enhancing}. In addition, the method's sensitivity to the initial conditions of the optimization problem can lead to poor or sub-optimal results if improper initial conditions are provided. In the next section, we will discuss a more efficient model for reconstructing a discrete dynamic object. 

%%%%%%%%%%%%%%%%%%%%%%%%%%%%%%%%%%%%%%%%%%%%%%%%%%%%%%%%%%%%%%%%%
%% Compressed Shape Sensing
%%%%%%%%%%%%%%%%%%%%%%%%%%%%%%%%%%%%%%%%%%%%%%%%%%%%%%%%%%%%%%%%%
\subsection{Compressed Shape Sensing}
\label{sec:Methods:CSS}
Compressed shape sensing (CSS) is an imaging method that directly encodes the discrete nature of the object under investigation. It combines the level-set representation of the imaging object, which enables accurate and efficient estimation of the object's shape and internal properties\cite{osher2004level, dorn2006level, santosa1996level}, with the benefits of compressive sensing. This combination minimizes the number of measurements required for image reconstruction, resulting in improved computational efficiency, reduced measurement time, and enhanced image quality\cite{kolehmainen2008limited,hamalainen2013sparse}.

%--------------------------------------------------------------------------------
\begin{figure}[t]
    \centering
    \includegraphics[width=0.6\textwidth]{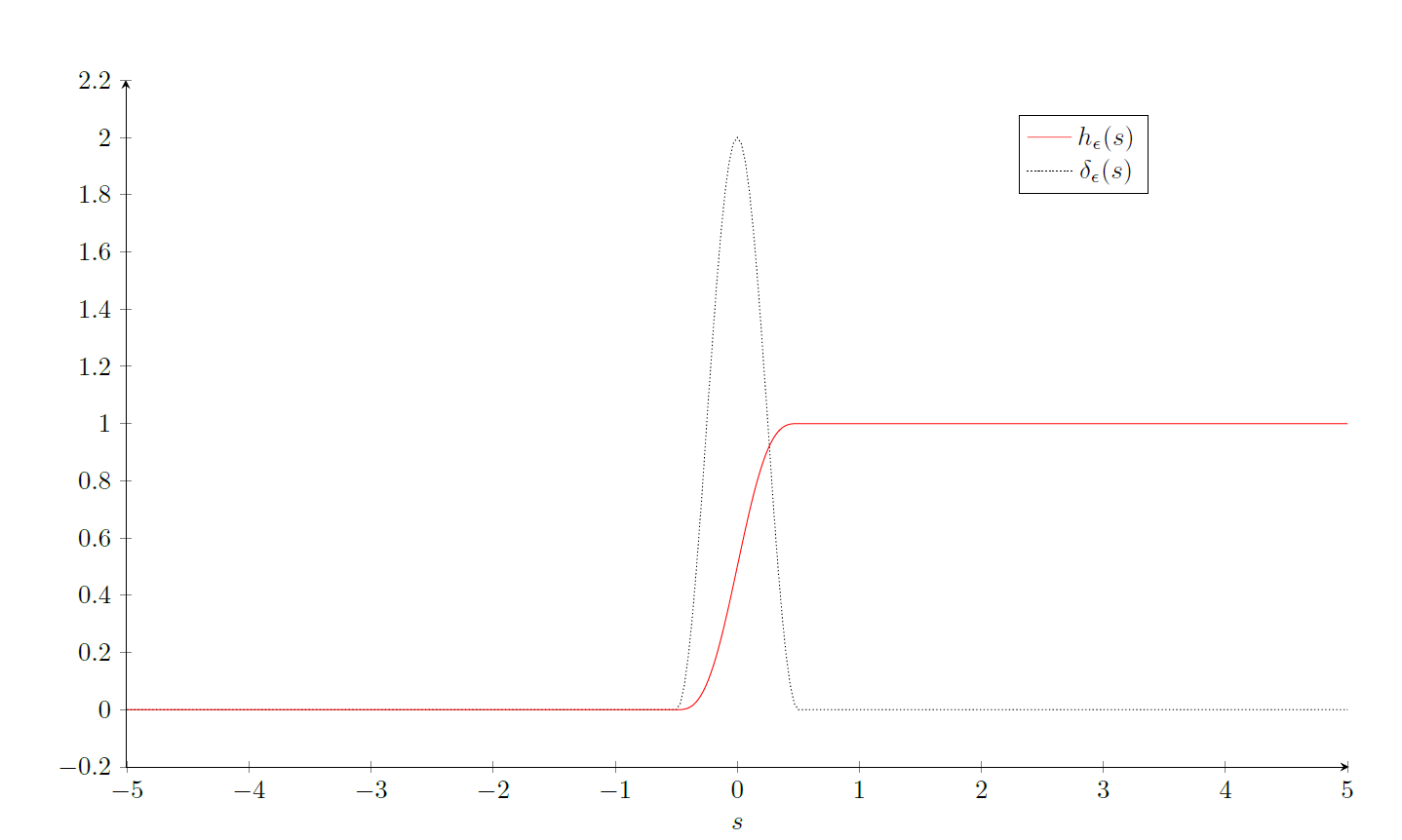}
    \caption{Smooth approximation of the Heaviside and it's derivative (which is a Dirac-Delta function).}
    \label{fig:smooth_heaviside}
\end{figure}
%--------------------------------------------------------------------------------

The level-set method represents an object's shape as the zero level set of a higher-dimensional function, known as the level-set function\cite{osher1988fronts}. The level set is defined as the set of all points $\mathbf{r}$ in the domain $\Omega \subset \mathbb{R}^d$ such that $\phi(\mathbf{r}) = 0$. A binary image function with support $\Gamma \subset \Omega$ can be modeled using the level-set method as $\vx(\mathbf{r}) = h(\phi(\mathbf{r}))$, where $h: \mathbb{R} \rightarrow \{0, 1\} $ is the Heaviside function and $\phi(\mathbf{r}) \geq 0$ if $\mathbf{r} \in \Gamma$ and $\phi(\mathbf{r}) < 0$ if $\mathbf{r} \notin \Gamma$. Since the level-set function still entails the same dimensionality as the image, it can represented in a chosen basis as
\begin{align*}
    \phi(\vr) = \sum_{i=1}^{k} \psi_i(\vr) \alpha_i,
\end{align*}
where $\psi_i: \mathbb{R}^d \rightarrow \mathbb{R}$ for $i =1, \dots, k$ are the basis functions, and $\valpha = \{\alpha_1, \dots, \alpha_k \}$ are the coefficients. For objects with smooth boundaries, the number of basis functions is much smaller than the dimensionality of the image, leading to the compression of the level-set function. This approach is also known as the parametric level-set method in the literature\cite{aghasi2011parametric, liu2017parametric, kadu2016salt, pingen2010parametric}. Inserting this ansatz into \eqref{eq:LeastSquares} results in
\begin{align}
    \vx_t^\star =  h (\Psi \valpha^\star_t ), \quad	\valpha_t^\star \in \underset{\| \valpha_t \|_1 \leq \tau  }{\argmin} \quad \| \mathcal{A}_t \! \left( h (\Psi \valpha_t ) \right) - \vy_t \|^2 
    \label{eq:CSS}
\end{align}
for each $t$: We found a binary-valued solution to \eqref{eq:FwdModel} by solving a real-valued optimization problem formulated using the level-set method. However, the Heaviside function, which is used to convert the binary problem to a real-valued problem, is non-continuous and its derivative is singular. Therefore, to make the objective differentiable with respect to the level-set function, a smooth approximation of the Heaviside function will be used:
\begin{align*}
    h_\epsilon (s) = \begin{cases}
    0 & \quad s \leq -\epsilon \\
    \frac{1}{2} \left(1 + \frac{s}{\epsilon} + \frac{1}{\pi} \sin \left( \frac{\pi s}{\epsilon} \right) \right) & \quad |s| < \epsilon \\
    1 & \quad s \geq \epsilon 
    \end{cases},
\end{align*}
where $\epsilon$ determines the width of the transition region between the two constant values (0 and 1) of the approximate Heaviside function. A smaller value of $\epsilon$ results in a sharper transition between 0 and 1, making the approximation closer to the actual Heaviside function. On the other hand, a larger $\epsilon$ value leads to a smoother transition between the two constant values, which can be beneficial when working with optimization algorithms that require differentiable functions. Figure~\ref{fig:smooth_heaviside} plots the approximation of the Heaviside function. This approximation permits us to employ conventional optimization methods suited for differentiable functions. However, it is important to highlight that the optimization task remains challenging due to its non-convex nature. In non-convex problems, several local minima may exist that do not represent the best possible outcome. Relying on local descent strategies might cause the optimization to settle at one of these suboptimal points. This complexity is a recognized challenge in the realm of optimization. Hence, while our refined function aids in utilizing certain optimization techniques, careful consideration is required in choosing initial points and optimization methods. Our objective with the continuous approximation is to efficiently find a good solution, even if confirming its optimality remains elusive.

The compressed shape sensing approach can be integrated in the binned reconstruction framework. However, binned reconstruction methods, which are decoupled over time $t$, have limited applicability in single-shot imaging scenarios. Hence, in the next subsection, we discuss the main contribution of this work, the dynamic extension of compressed shape sensing approach.

%%%%%%%%%%%%%%%%%%%%%%%%%%%%%%%%%%%%%%%%%%%%%%%%%%%%%%%%%%%%%%%%%
%% Dynamic Shape Sensing
%%%%%%%%%%%%%%%%%%%%%%%%%%%%%%%%%%%%%%%%%%%%%%%%%%%%%%%%%%%%%%%%%
\subsection{Dynamic Shape Sensing}
\label{sec:Methods:DSS}
The level-set method can be trivially extended to dynamic shapes letting the level-set function evolve in time, $\phi: \Omega \times [0,T] \to \mathbb{R}$ \cite{niemi2015dynamic, haario2017shape}.
Similar to the compressed shape sensing approach, the fundamental principle of spatiotemporal compression of level-set function is to represent it using a set of basis functions. In this work, we use the discrete cosine basis to compress the level-set function. The Discrete Cosine Transform (DCT) is a mathematical tool that is utilized to decompose a signal into its harmonic components, which can be used for variety of tasks, including, image compression, signal processing, and motion analysis.

The DCT can be mathematically formulated as follows: for a level-set function $\phi(\vr, t)$, where $\vr$ represents the spatial coordinates and $t$ represents the time index, the DCT can be used to represent the spatiotemporal level-set function using the DCT coefficients $\alpha(\vw, s)$, where $\vw \in \R^d$ represents the spatial frequencies, and $s \in \R$ represents temporal frequency:
\begin{align*}
    \phi(\vr,t) = \sum_{\vw,s} \alpha(\vw, s) \psi(\vr,t,\vw, s) ,
\end{align*}
where the $\psi: \R^d \times \R \times \R^d \times \R \mapsto \R$ is a DCT kernel with following formula
\begin{align*}
    \psi(\vr,t,\vw, s) =  \prod_{i=1}^{d} \cos \left( \frac{\pi r_i (2 w_i +1)}{2 W_i } \right) \cos\left( \frac{\pi t(2 s +1)}{2T} \right) ,
\end{align*}
where $W_i$ represents the total number of spatial frequencies in the $i^{\text{th}}$ direction. This representation is commonly known as DCT-II transform in the literature\cite{rao1990discrete}. The discretization of the DCT results in the following linear algebraic relationship:
\begin{align*}
    \vphi &= \Psi \valpha,
\end{align*}
where $\Psi$ denotes the kernel, and $\valpha$ denotes the DCT coefficients.

The utilization of a reduced set of DCT coefficients to represent the level-set function that characterizes the motion of discrete objects can be achieved through the assumption of smooth motion of the objects under examination. Given that single-shot dynamic imaging instruments are capable of achieving maximum possible temporal resolution, it is natural to assume that the motion of the objects will be smooth. As a result, the level-set function describing such smooth motion will also exhibit smooth characteristics and, to a certain extent, exhibit lower dimensionality. Consequently, it is feasible to restrict the DCT coefficients to a subset of $k$ components, thereby enabling the compression of the temporal images representation.

\begin{figure}[!t]
    \centering
    \includegraphics[width=0.5\textwidth]{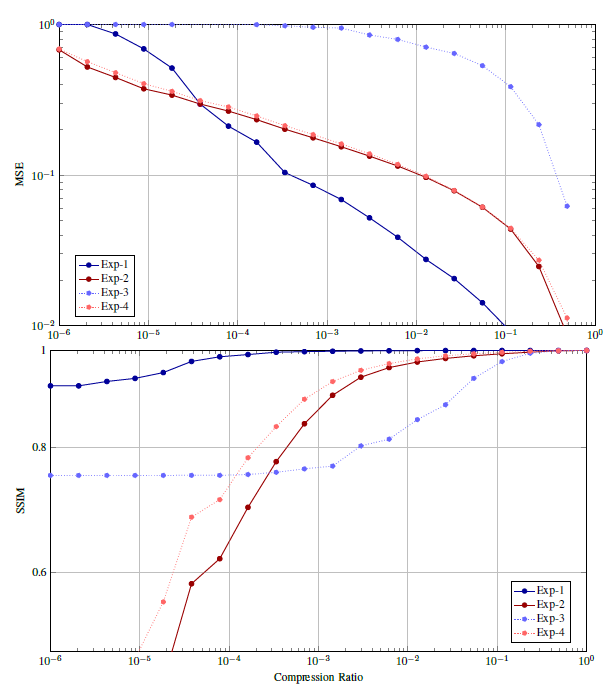}
    \caption{Comparison of compression techniques for smooth (Exp 1-2) and randomized (Exp 3-4) spatiotemporal volumes using MSE and SSIM metrics. Exp 1 and 3 employ dynamic shape sensing with DCT-based level-set function compression, while Exp 2 and 4 use direct DCT-based compression in the image domain. The smooth spatiotemporal volume is shown in Figure~\ref{fig:phantoms}(a), while the randomized volume is created using the random permutation in time-axis of the same phantom.}
    \label{fig:compression}
\end{figure}

To illustrate this core concept of our technique, we prepared an example in Figure~\ref{fig:compression}, which demonstrates the impact of assuming smooth object motion and using a Heaviside function for level-set compression in temporal image compression. The figure presents a comparison of four experiments, with Experiments 1 and 2 operating under smooth motion assumptions, while Experiments 3 and 4 involve randomized motion. More specifically, Experiments 1 and 3 employ DCT-based level-set function compression using a Heaviside function, while Experiments 2 and 4 utilize direct DCT-based compression of the spatiotemporal volume. The top graph, which maps mean-squared error (MSE) against the compression ratio, reveals a universal decrease in MSE as compression increases for all experiments. However, Experiments 1 and 2, under smooth motion assumptions, demonstrate a faster rate of decrease. Remarkably, Experiment 1, which incorporates the Heaviside function, shows the fastest convergence, highlighting the combined efficiency of smooth motion assumptions and the Heaviside function in reducing errors. Similarly, the bottom graph, which plots the structural similarity index (SSIM) against the compression ratio, shows that despite varying levels of compression, Experiments 1 and 2 maintain consistently high SSIM values. Notably, Experiment 1, employing the Heaviside function under smooth motion, outperforms the others, reinforcing the capacity of our proposed method to preserve image structure under compression.

%--------------------------------------------------------------------------------
%--------------------------------------------------------------------------------

It is worth noting that this is just one example of a dynamic shape sensing model, other types of basis functions and different sparsity-promoting terms could be used, depending on the characteristics of the measurements and the requirements of the specific application. For instance, the use of wavelet basis functions, such as the discrete wavelet transform, can provide a more localized representation of the motion information.

The resulting formulation, termed as \textit{dynamic shape sensing} (DSS), is posed as a least-squares problem, where the goal is to find the coefficients of the DCT of the level-set function of the object from the time-series of measurements:
\begin{equation}
    \valpha^\star \in \underset{\| \valpha \|_1 \leq \tau }{\argmin} \quad \sum_{t=1}^{T} \| \mathcal{A}_t \! \left( h_\epsilon \! \left( \Psi_t \valpha \right) \right) - \vy_t \|^2 
    \label{eq:DSS}
\end{equation}
The solution, \ie , the series of images, can be obtained by applying the Heaviside function to the optimized DCT coefficients:
\[
	\vx_t^\star = h \left(  \Psi_t \valpha^\star \right) \qquad \forall \, t=1, \dots, T
\]
We describe the extensions of this framework to various scenarios in Appendix~\ref{sec:Extensions}.

%%%%%%%%%%%%%%%%%%%%%%%%%%%%%%%%%%%%%%%%%%%%%%%%%%%%%%%%%%%%%%%%%
%% Optimization
%%%%%%%%%%%%%%%%%%%%%%%%%%%%%%%%%%%%%%%%%%%%%%%%%%%%%%%%%%%%%%%%%
\subsection{Optimization Strategies}
\label{sec:Methods:Optimization}

To solve the dynamic shape sensing problem, we can make use of an proximal iterative scheme, \eg, by using the gradient of the loss function:
\begin{align*}
    \valpha^{(p+1)} \gets \mathcal{P}_{\tau} \left( \valpha^{(p)} - \gamma \nabla \mathcal{J} \! \left( \valpha^{(p)} \right) \right)
    % \valpha \leftarrow \mathcal{P}_{\tau} \big( \valpha - \gamma \nabla \mathcal{J} \! \left( \valpha \right) \big) 
\end{align*}
where $\mathcal{J}(\valpha) = \sum\nolimits_{t=1}^{T} \| \mathcal{A}_t \! \left( h_\epsilon \! \left( \Psi_t \valpha \right) \right) - \vy_t \|^2$ is the objective function. This method involves starting with an initial estimate of the shape parameter and iteratively updating the estimate through the use of a gradient descent step followed by projection onto constraints, as shown in the equation above. The step size, or learning rate, $\gamma$, is chosen through the use of a linesearch algorithm to ensure that the objective function is decreasing at each iteration. The term $\mathcal{P}_{\tau}$ denotes the projection onto the $\ell_1$ norm of size $\tau$. This projection acts as a regularizing step, constraining the coefficients $\valpha$ to a specific range defined by $\tau$. This constraint helps in producing more robust and sparse solutions, particularly beneficial when there's a need to extract meaningful features from the shape representations or when overfitting is a concern.

The gradient of the objective function with respect to the shape parameter, $\nabla \mathcal{J}(\valpha)$, can be computed using the gradient of the objective function with respect to the state variable, $\nabla \mathcal{J}(\vx_t)$, and the orthonormal basis, $\Psi_t$:
\begin{align*}
	\nabla \mathcal{J}(\valpha) &=  \sum_{t=1}^{T} 2 \Psi_t^H \! \! \left( \diag \left(\delta_\epsilon(\Psi_t \valpha) \right) \mathcal{A}_t^H \! \! \left( \mathcal{A}_t \! \left( h_\epsilon \! \! \left( \Psi_t \valpha \right) \right) - \vy_t \right) \right) ,
\end{align*}
where $\diag$ represents the diagonal matrix, and $\delta_\epsilon$ is the derivative of $h_\epsilon$, an approximation of the dirac-delta function with width $\epsilon > 0$ that has following form:
\begin{align*}
    \delta_\epsilon (s) = \begin{cases}
    \frac{1}{2 \epsilon} \left(1 +  \cos \left( \frac{\pi s}{\epsilon} \right) \right) & \quad |s| \leq \epsilon \\[1ex]
    0 & \quad \text{otherwise} 
    \end{cases}.
\end{align*}
$\mathcal{A}_t^H$ and $\Psi_t^H$ are the adjoint operator for all $t=1, \dots, T$. Additionally, the adjoint of the orthonormal basis, $\Psi^H$, can be computed efficiently using the basis itself, \ie, 
\begin{align*}
    \Psi^H \! \! \left( \phi(\vr, t); \vw, s \right) = \sum_{\vr,t} \phi(\vr,t) \, \psi (\vr,t,\vw, s).
\end{align*}
Computationally, the inverse of the DCT-II can be computed using the DCT-III transform. The adjoint operator of various forward operators are given in Appendix~\ref{sec:Adjoint}.

One of the issue with gradient-based iterative minimization schemes for \eqref{eq:DSS} is that the hyperparameter $\epsilon$ controls the width of the Heaviside function, which affects the gradient at every iteration. A well-known issue with the level-set method is that the level-set function can become flat or steep, leading to poor convergence to the solution. To address this issue, we use the following heuristic scheme as presented in \cite{kadu2016salt}:
\begin{align*}
    \epsilon = \kappa \max \left( | \nabla \phi | \right),
\end{align*}
where $\kappa$ is initialized to $0.1$. The DSS Algorithm, outlined in Algorithm~\ref{alg:DSS}, refines an object's shape estimate using dynamic tomographic measurements. Starting with an initial shape and its DCT coefficients, the algorithm employs two nested loops: the outer loop adjusts the width parameter $\kappa$ of the approximated Heaviside function, making the approximation progressively closer to the true Heaviside function, while the inner loop optimizes the DCT coefficients to minimize the loss of the objective function, ensuring they remain within the $\ell_1$-norm ball constraints.  A step size is determined using backtracking scheme\cite{nocedal1999numerical} while the projection is carried out using fast simplex algorithm\cite{condat2016fast}. After completing both sets of iterations, the final shape estimate is derived by applying the true Heaviside function to the updated DCT coefficients. This structure ensures an efficient and robust shape estimation.

%--------------------------------------------------------------------------------
\begin{algorithm}[!t]
\caption{Dynamic Shape Sensing Algorithm}
\begin{algorithmic}[1]
\Require{operators $\mathcal{A}_t$, measurements $\vy_t$ for $t = 1, \dots, T$}
\Require{Regularization parameter $\tau$, DCT dictionary $\Psi$}
\Require{Number of inner and outer iterations $M$, $N$}
\Ensure{$\mathbf{X}^{\star}$}
\Statex
\State Compute initial estimate $\mathbf{X}$
\State Compute the initial estimate of DCT coefficients $\boldsymbol{\alpha}$ from $\mathbf{X}$
\State Initialize the Heaviside width with $\kappa = 0.1$
\For{$q=1, \dots, M$}
    \For{$p=1, \dots, N$}
        \State Compute the gradient: $\vg \triangleq \nabla \mathcal{J}\! \left(\boldsymbol{\alpha}\right)$
        \State Compute the step size: $\gamma_p$
        \State Compute the next iterate: $\boldsymbol{\alpha} = \mathcal{P}_{\tau} \!  \left(\boldsymbol{\alpha} - \gamma_p \vg \right)$
    \EndFor
    \State $\kappa \leftarrow 0.8 \kappa$
\EndFor
\State Compute the final solution $\mathbf{X}^{\star} = h(\Psi \boldsymbol{\alpha})$
\end{algorithmic}
\label{alg:DSS}
\end{algorithm}
%--------------------------------------------------------------------------------

% \input{sections/experiments}
%%%%%%%%%%%%%%%%%%%%%%%%%%%%%%%%%%%%%%%%%%%%%%%%%%%%%%%%%%%%%%%%%
%% Numerical Experiments
%%%%%%%%%%%%%%%%%%%%%%%%%%%%%%%%%%%%%%%%%%%%%%%%%%%%%%%%%%%%%%%%%
\section{Numerical Experiments}
\label{sec:numExp}

%----------------------------------------------------------------
\begin{figure*}[!htb]
    \centering
    \begin{tabular}{c|c|c}
        \includegraphics[width=0.3\textwidth]{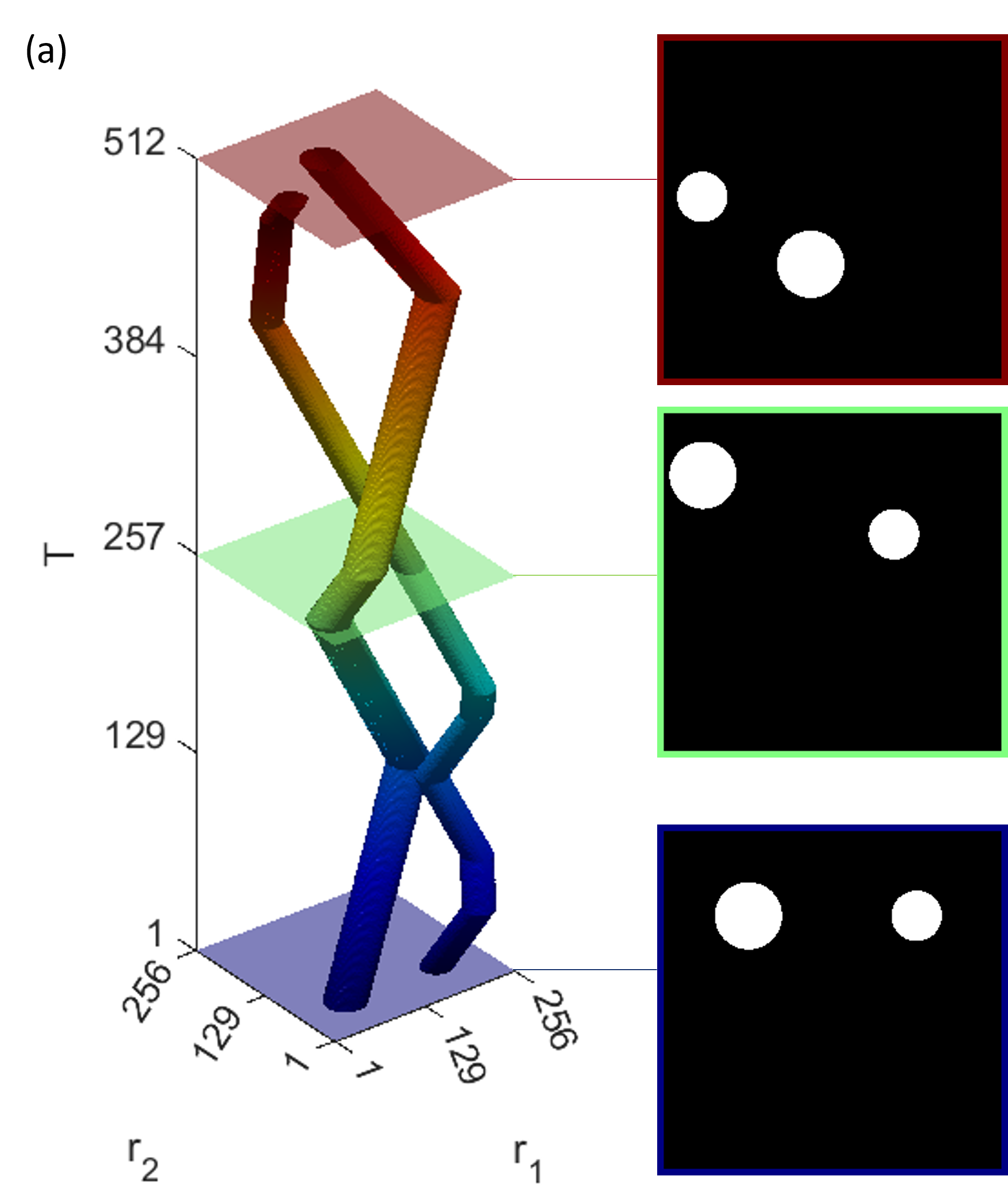} & \includegraphics[width=0.3\textwidth]{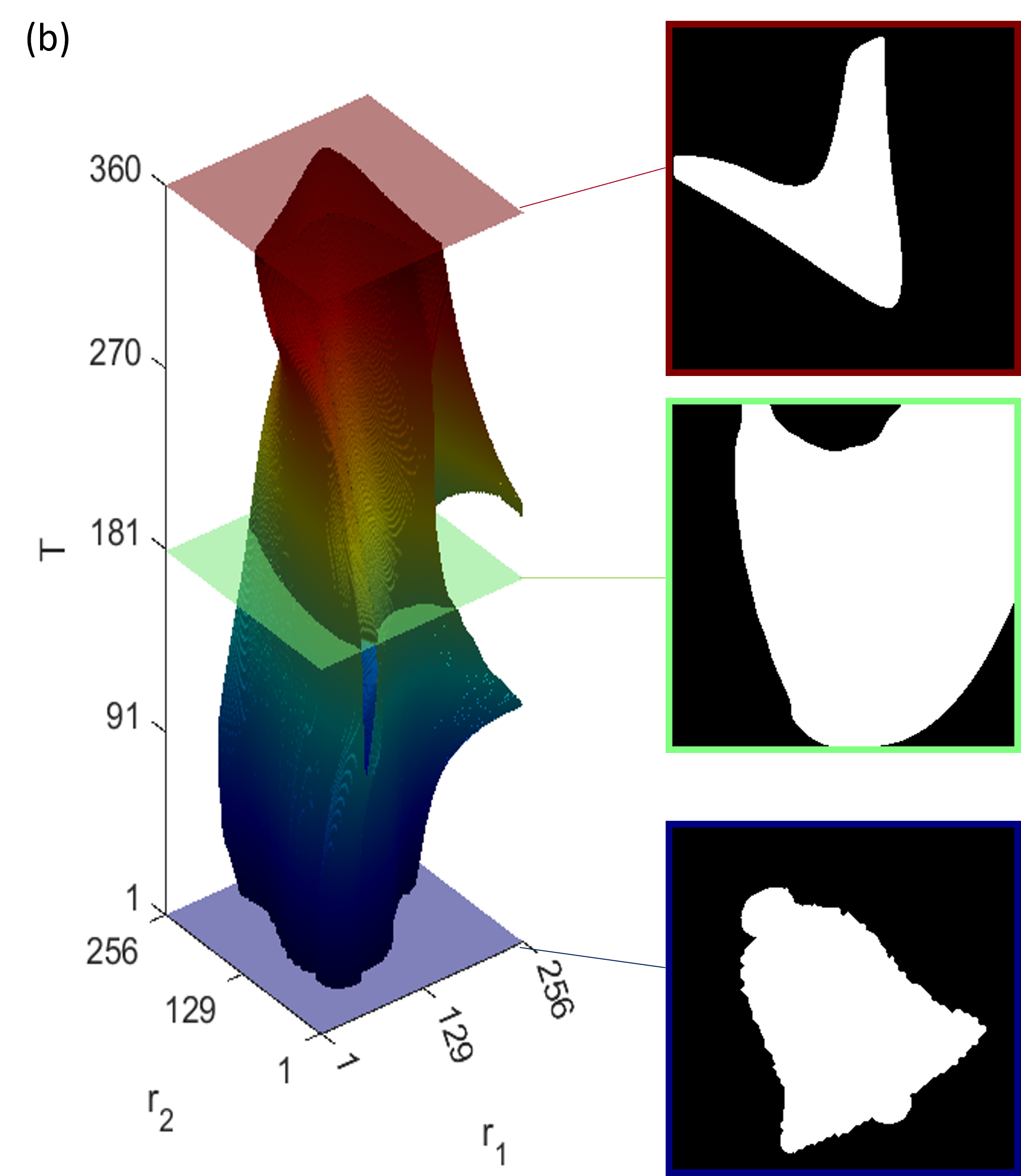} & \includegraphics[width=0.3\textwidth]{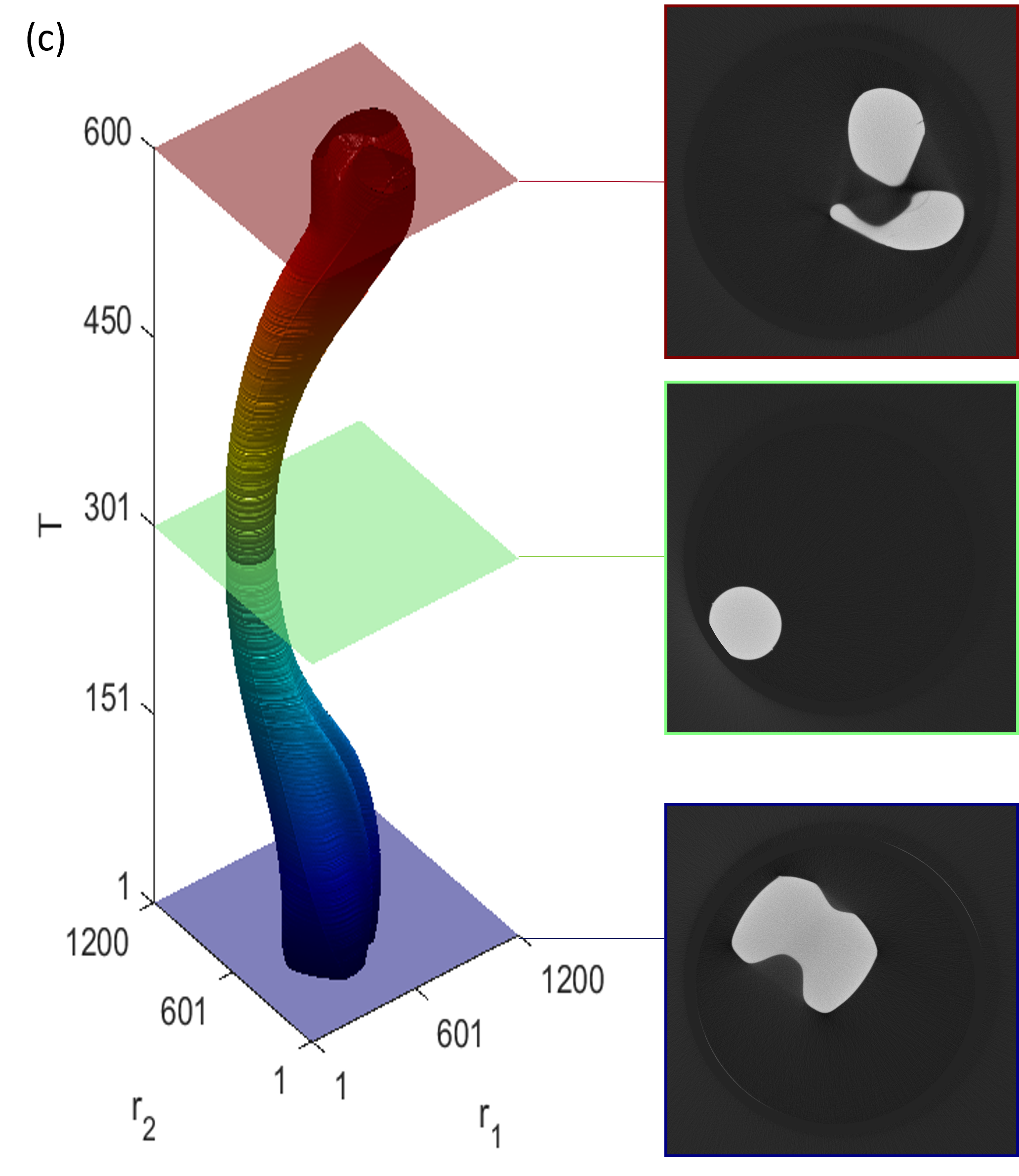}
    \end{tabular}
    \caption{Time-lapse of (a) rigid motion phantom with three time-slices at $t = {1, 256, 512}$, (b) non-rigid motion phantom with three time-slices at $t = {1, 180, 360}$, and (c) DogToy experiment with three time-slices at $t = {1, 300, 600}$.}
    \label{fig:phantoms}
\end{figure*}
%----------------------------------------------------------------
\begin{figure*}[!htb]
    \centering
    \includegraphics[width=0.9\textwidth]{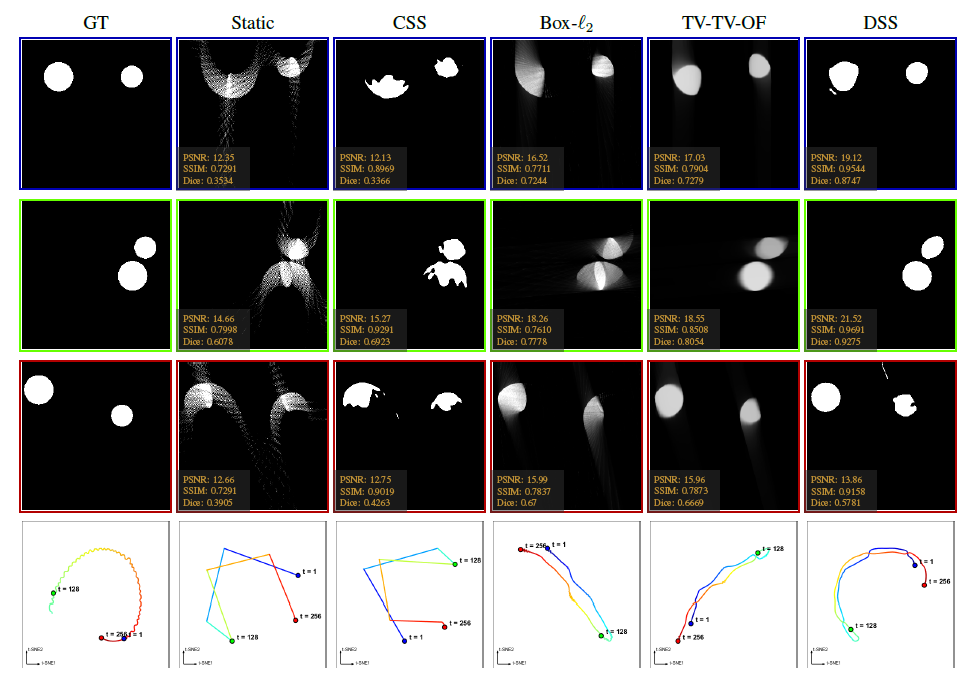}
    \caption{Comparison of reconstruction algorithms for synthetic ball motion experiment: temporal slices (top three rows) display Ground Truth (GT), Static, CSS, Box-$\ell_2$, TV-TV-OF, and DSS results with corresponding PSNR, SSIM, and Dice scores overlaid. Higher scores indicate better reconstruction. The t-SNE plot (last row) visualizes the performance of each algorithm in the latent space; frames from consistent reconstructions cluster closer together, aiding in evaluating algorithmic distinction and similarity.}
    \label{fig:Ball:CT}
\end{figure*}
%----------------------------------------------------------------

%%%%%%%%%%%%%%%%%%%%%%%%%%%%%%%%%%%%%%%%%%%%%%%%%%%%%%%%%%%%%%%%%
%% Implementation
%%%%%%%%%%%%%%%%%%%%%%%%%%%%%%%%%%%%%%%%%%%%%%%%%%%%%%%%%%%%%%%%%
\subsection{Implementation} 
\label{sec:NumExp:Implementation}
Our proposed method and all comparison methods were implemented using MATLAB and will be made available on GitHub upon the paper's acceptance. The main operations include the forward and inverse modeling of the tomographic imaging process and the optimization procedure for the reconstruction of the dynamic images. The forward and inverse models were simulated with the ASTRA toolbox \cite{bleichrodt2016easy,van2016fast}, using parallel-beam geometry and adding additive white Gaussian noise (AWGN) to the measurements to reduce inverse crime effects.

For various reconstruction approaches, such as static, CSS, and Box-$\ell_2$ regularized, we employed the spectral projected gradient  scheme\cite{schmidt2009optimizing}, limiting the procedure to a maximum of 1000 iterations. For the DSS approach, we employed Algorithm~\ref{alg:DSS}, setting parameters as $M=20$, $N=30$. The basis $\Psi$ was defined using DCT-II, retaining only the top $1\%$ coefficients in each dimension. In contrast, the alternating direction method of multipliers (ADMM) \cite{BoPaChPeEc11,lucka2018enhancing} was used to solve the convex optimization sub-problems for the TV-TV-OF method. All computations were executed on an computing server with a $3^\text{rd}$ Gen. AMD CPU with 16 cores and 1TB RAM.

We used the peak signal-to-noise ratio (PSNR), structural similarity index (SSIM)\cite{wang2004image}, and Dice coefficient (Dice)\cite{dice1945measures} to evaluate the reconstruction results for both individual frames and the overall sequences. Higher PSNR, SSIM, and Dice values indicate better reconstruction accuracy. The Dice coefficient was calculated on the binarization of the reconstruction results, which was achieved using Otsu's thresholding algorithm\cite{otsu1979threshold}. To provide insights into the spatio-temporal relationships among the reconstructed frames, we utilized \textit{t-distributed Stochastic Neighbor Embedding} (t-SNE) on the reshaped time-resolved data\cite{van2008visualizing}. The t-SNE plots present these data in 2D, where similar temporal frames cluster closer together. This representation facilitates a clear distinction of the temporal evolution captured by each algorithm. The learning rate was set to 500, with a perplexity of 50, and an exaggeration factor of 10. The maximum allowed iterations were set to $10^6$ with a tolerance of $10^{-10}$.

%%%%%%%%%%%%%%%%%%%%%%%%%%%%%%%%%%%%%%%%%%%%%%%%%%%%%%%%%%%%%%%%%
%% Synthetic Experiments
%%%%%%%%%%%%%%%%%%%%%%%%%%%%%%%%%%%%%%%%%%%%%%%%%%%%%%%%%%%%%%%%%
\subsection{Synthetic Experiments}

Our synthetic experiments were conducted on two numerical phantoms: a rigid motion of two disks and a non-rigid deformation of a Bell-shaped phantom. Both phantoms were discretized at a resolution of $512 \times 512$ pixels. We began with a starting angle of $\vtheta_1 = 0^\circ$, with an angular difference of $\delta_{\vtheta} = 5^\circ$ between consecutive frames. We compared the performance of our proposed method DSS with four other methods: static reconstructions, CSS, Box-$\ell_2$ regularized reconstruction, and TV-TV-OF. We noted that the first two methods use binning in the temporal direction, while the remaining two do not.

%----------------------------------------------------------------
\begin{figure*}[!htb]
    \centering
    \includegraphics[width=0.9\textwidth]{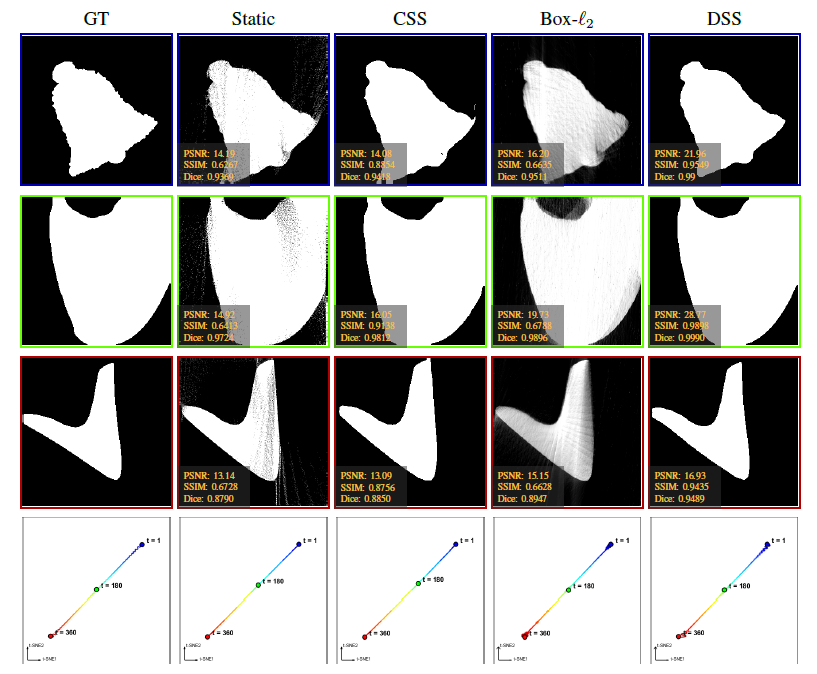}
    \caption{Comparison of reconstruction algorithms for non-rigid motion experiment: Temporal slices (top three rows) and t-SNE plot (last row) representing the performance of Ground Truth (GT), Static, CSS, Box-$\ell_2$, and DSS methods.}
    \label{fig:Bell:CT}
\end{figure*}
%----------------------------------------------------------------

%%%%%%%%%%%%%%%%%%%%%%%%%%%%%%%%%%%%%%%%%%%%%%%%%%%%%%%%%%%%%%%%%
%% Rigid motion phantom
%%%%%%%%%%%%%%%%%%%%%%%%%%%%%%%%%%%%%%%%%%%%%%%%%%%%%%%%%%%%%%%%%
\subsubsection{Rigid motion phantom} 
\begin{figure*}[!htb]
    \centering
    \includegraphics[width=0.9\textwidth]{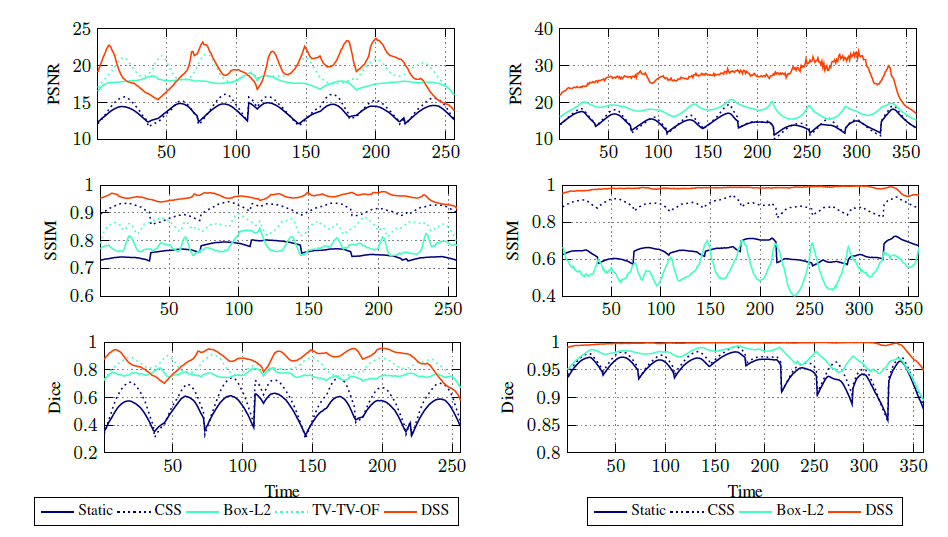}
    \caption{Reconstruction performance comparison for synthetic dynamic X-ray tomography of rigid (left) and non-rigid (right) phantoms using PSNR, SSIM, and Dice coefficient metrics across time.}
    \label{tab:BallBell:CT}
\end{figure*}

We generated a dynamic 2D spatiotemporal phantom simulation featuring two balls traveling at a constant speed within a 2D space over $512$ frames. We appropriately initialized the velocities and positions of the two balls to model the desired motion. The video generation process commenced by initializing the balls and a video frame with zeros. Subsequently, we executed a loop that updated the balls' positions, checked for collisions with the 2D space boundaries, and adjusted their velocities accordingly. Finally, we updated the balls' positions once more, and the video frame was refreshed with the new positions of the balls (Figure~\ref{fig:phantoms}). The results for 3 temporal slices are shown in Figure~\ref{fig:Ball:CT}. These images clearly indicate the necessity of considering the temporal dimension in the reconstruction algorithm due to the shortcomings of static and CSS methods. While the Box-$\ell_2$ reconstruction results outperformed static and CSS methods, they still exhibited prominent motion imprints. The TV-TV-OF and DSS methods demonstrated minimal motion imprints and offered more precise reconstruction results. This is further substantiated by the PSNR, SSIM, and Dice coefficients as shown in Figure~\ref{tab:BallBell:CT}.

%----------------------------------------------------------------
%----------------------------------------------------------------
%----------------------------------------------------------------
\begin{figure*}[!htb]
    \centering
    \includegraphics[width=0.9\textwidth]{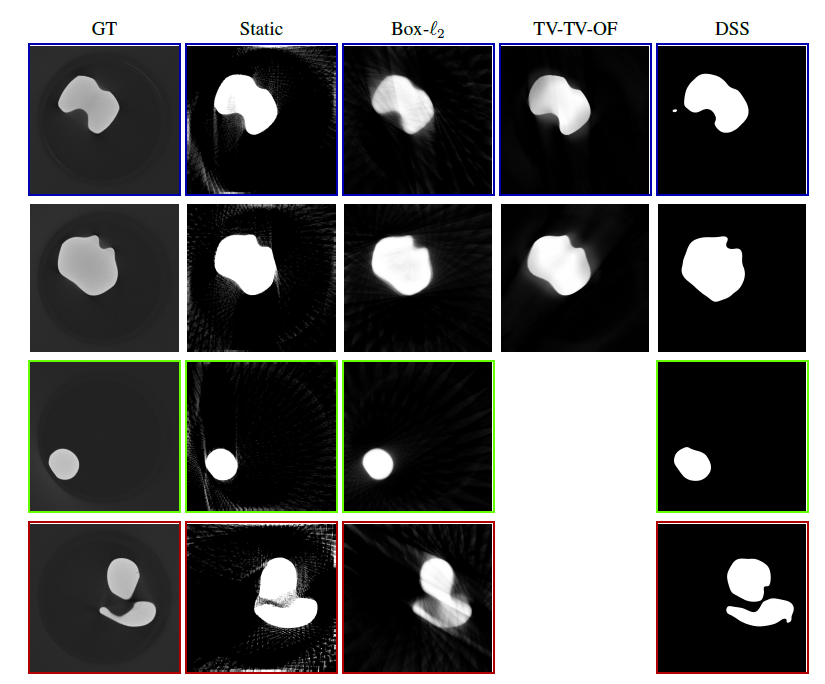}
    \caption{Comparative Analysis of Reconstruction Algorithms in the DogToy Experiment: Temporal slices showcasing the performance across methods—Ground Truth (GT), static, box-$\ell_2$, TV-TV-OF, and DSS. Specifically, the second row depicts results from temporal slice 64, while other corresponds to slices shown in Figure~\ref{fig:phantoms}. Note that the TV-TV-OF method was only run on 64 temporal slices due to computational limitations.}
    \label{fig:DogBone}
\end{figure*}
%----------------------------------------------------------------

%%%%%%%%%%%%%%%%%%%%%%%%%%%%%%%%%%%%%%%%%%%%%%%%%%%%%%%%%%%%%%%%%
%% Non-rigid motion
%%%%%%%%%%%%%%%%%%%%%%%%%%%%%%%%%%%%%%%%%%%%%%%%%%%%%%%%%%%%%%%%%
\subsubsection{Non-rigid motion phantom}
We generated a dynamic phantom consisting of a sequence of binary images that exhibit non-rigid deformation. To create the phantom, we initially imported an input image of a Bell phantom and resized it to $512 \times 512$ pixels. Next, we produced a sequence of meshes with sinusoidal deformation patterns at a default frequency of $10$ and amplitude of $2$ across $360$ temporal frames. We computed optical flow between the input image and the warped image at each time step using Lucas-Kanade algorithm\cite{lucas1981iterative}. The mesh was then updated based on the flow, employing the iterative closest point (ICP) method\cite{besl1992method}. A binary image was generated from the warped image, and morphological operations like erosion, dilation, and filling were applied to eliminate holes and smooth the binary image. The resulting phantom was a 3D array with dimensions of $512 \times 512 \times 360$ (Figure~\ref{fig:phantoms}).

The reconstruction results for 3 temporal slices are illustrated in Figure~\ref{fig:Bell:CT} (we did not compute the TV-TV-OF reconstruction here, as the motion model it employs is not suitable for non-rigid deformation). Noteably, the static and CSS algorithms do not exhibit significant motion imprints, which were observed in the rigid motion phantom. This could be attributed to the fact that the non-rigid phantom does not involve translation or rotation, which are the most challenging types of motion for static and CSS methods. Nevertheless, some areas of the reconstructed images remain ambiguous regarding the presence of the object, highlighting the limitations of these methods. This is evident in the poor performance concerning PSNR, SSIM, and Dice coefficient, as presented in Figure~\ref{tab:BallBell:CT}. The Box-$\ell_2$ method does not enhance the results compared to static or CSS methods, likely due to inaccuracies of the underlying motion model. Conversely, the DSS method accurately reconstructs the motion in this phantom, as demonstrated in Figure~\ref{fig:Bell:CT} and Figure~\ref{tab:BallBell:CT}. The DSS method models the non-rigid deformation of the shape using a set of basis functions, enabling it to capture the complex and non-linear motion in the phantom. The accuracy of the DSS method in this case suggests its potential as a promising approach for dynamic tomographic imaging involving non-rigid motion.

%%%%%%%%%%%%%%%%%%%%%%%%%%%%%%%%%%%%%%%%%%%%%%%%%%%%%%%%%%%%%%%%%
%% Real Dataset
%%%%%%%%%%%%%%%%%%%%%%%%%%%%%%%%%%%%%%%%%%%%%%%%%%%%%%%%%%%%%%%%%
\subsection{Real-world datasets}
To evaluate the performance of our method in realistic scenarios, we acquired a pseudo-dynamic 2D dataset at the FleX-ray lab of CWI\cite{CoLuPaLoBa20}, which hosts a laboratory cone-beam CT scanner. The object consists of a dog toy in the shape of a bone, compressed inside a cylindrical cardboard tube. The term ``pseudo-dynamic" denotes considering the third dimension of the dog toy sample as time. We acquired 673 time-frames (i.e., z-slices). For each 2D slice, we captured X-ray projections at 1200 equidistant angles in $[0, 2\pi]$ and read out the central line of the flat panel detector, which consists of 956 pixels at \SI{149.6}{\micro\metre^2} pixel size.  The X-ray tube voltage was \SI{90}{\kilo\volt} and a copper filter was used to bloc the low-energy part of the spectrum to limit beam-hardening artifacts. The source-to-detector distance was \SI{487.9}{\milli\meter}, while the source-to-origin of the sample was \SI{374.5}{\milli\meter} in a fan-beam geometry. For the computations, we employed an image spatial grid of $1200 \times 1200$ pixels. To generate a ground-truth image, we applied the filtered backprojection algorithm, resulting in the images displayed in the leftmost column of Figure~\ref{fig:DogBone}.

This pseudo-dynamic dataset simulates a non-rigid motion. We selected an angle difference of $\delta_{\theta} = 5^\circ$ to convert the dataset into a single-shot dynamic imaging problem. We tested static, CSS, Box-$\ell_2$, TV-TV-OF and DSS algorithms on this dataset. As running TV-TV-OF on the complete dataset required memory exceeding 1TB and computational times exceeding one week, we limited this method to the first $64$ frames, only. Figure~\ref{fig:DogBone} illustrates the reconstruction of four temporal frames. Given that the presumed ground truth incorporates beam hardening artefacts, a quantitative comparison is not presented. Our findings indicate that temporal regularization enhances reconstruction accuracy over other techniques. Notably, DSS outperforms its counterparts based on visual inspection.

% \input{sections/discussions}
%%%%%%%%%%%%%%%%%%%%%%%%%%%%%%%%%%%%%%%%%%%%%%%%%%%%%%%%%%%%%%%%%
%% Discussions
%%%%%%%%%%%%%%%%%%%%%%%%%%%%%%%%%%%%%%%%%%%%%%%%%%%%%%%%%%%%%%%%%
\section{Discussion}
\label{sec:Discussion}

Our experiments show that using advanced image models that take into account both spacial and temporal characteristics can improve image quality in dynamic single-shot tomographic imaging of discrete objects and suppress motion artifacts in particular when the motion is moderate or severe. However, while the TV-TV-OF method outperforms simpler variational regularization schemes like static or Box-$\ell_2$, it requires considerably more computational time and resources and has more hyper-parameters to tune. The DSS method can reconstruct a accurate, discrete 3D volume over time with less computational time and resources. However, it's important to remember that DSS assumes that the movement of objects can be represented accurately by a small set of predefined functions. This might not always be true, especially when the motion is complicated or doesn't follow a regular pattern. Choosing the right basis functions for DSS requires to careful balance accuracy and computational complexity. Some possible choices include the discrete cosine transform, the Haar wavelet transform, and the Legendre polynomials. Our findings suggest that other techniques, like TV-TV-OF, could still be useful in situations where their strengths outweigh their weaknesses, even if DSS generally performs better.

%%%%%%%%%%%%%%%%%%%%%%%%%%%%%%%%%%%%%%%%%%%%%%%%%%%%%%%%%%%%%%%%%
%% Conclusions
%%%%%%%%%%%%%%%%%%%%%%%%%%%%%%%%%%%%%%%%%%%%%%%%%%%%%%%%%%%%%%%%%
\section{Conclusion}
\label{sec:Conclusion}

The problem of image reconstruction of discrete objects from single-shot dynamic tomographic imaging poses significant challenges. However, our \textit{dynamic shape sensing} framework presents an innovative approach that integrates spatial and temporal smoothness of discrete objects into the reconstruction process and employs an iterative optimization algorithm to efficiently solve the resulting optimization problem. Our study's findings contribute to the ongoing discourse in the field, highlighting the potential of DSS in overcoming some of the challenges inherent in dynamic tomographic imaging. Furthermore, this work underscores the importance of adopting a more comprehensive spatiotemporal model to handle motion artefacts effectively, illuminating the limitations of conventional regularization-based methods.

Our future work involves refining the DSS method by testing it in diverse dynamic imaging situations with real-world datasets. We aim to integrate a stochastic gradient approach to boost efficiency and minimize computational effort. Additionally, we plan to improve the reconstruction quality by including physical models and machine learning techniques. We foresee these steps to broaden the application of the DSS method and improve its effectiveness significantly.

% appendices
\appendices
%%%%%%%%%%%%%%%%%%%%%%%%%%%%%%%%%%%%%%%%%%%%%%%%%%%%%%%%%%%%%%%%%%%
%%         FORWARD MODELS
%%%%%%%%%%%%%%%%%%%%%%%%%%%%%%%%%%%%%%%%%%%%%%%%%%%%%%%%%%%%%%%%%%%
\section{Forward Models}
\label{sec:ForwardModels}
The forward models in different imaging modalities provide the theoretical basis for the transformation from the object's physical properties to the collected data.

In dynamic Magnetic Resonance Imaging, the forward model describes the relationship between the object's magnetic susceptibility and relaxation properties and the measured data at each time point. This forward model is a linear operator, referred to as the Fourier Transform Model (FTM) \cite{fessler2010model,plewes2012physics}. It can be described mathematically as:
\begin{align*}
    \mathcal{A}_t(x(\vr, t); \vk \in K_t) = \int_{-\infty}^{\infty} x(\vr,t) \exp(-2\pi i \left\langle \vk, \vr \right\rangle ) \mathrm{d} \vr, 
\end{align*}
where $x(\vr, t)$ represents the object's magnetic susceptibility and relaxation properties in the spatial domain at time $t$, $\vk \in K_t$ represents the spatial frequencies in the k-space domain with $K_t$ being the set of frequencies (\eg, frequencies along a radial line as shown in Figure~\ref{fig:single_shot_exp}) acquired at time $t$. 

In ultrasound computed tomography, one forward model consists of the wave equation describing the propagation of ultrasonic waves through a medium with an unknown sound speed distribution $x(\vr,t)$:
\begin{align*}
    \frac{\partial^2 p(\vr,t)}{\partial t^2} - x(\vr,t)^2 \nabla^2 p(\vr,t) = -\frac{1}{\rho} q(\vr, t),
\end{align*}
where $p(\vr,t)$ denotes the pressure at a location $\vr$ at time $t$, $q(\vr,t)$ is the source term, and $\rho$ is the density of the medium, which is assumed to be constant (and known). The forward operator, $\mathcal{A}_t^{\text{US}}$, is defined as the non-linear mapping of $x(\vr,t)$ to the restriction of the solution of the wave equation $p(\vr,t)$ to a set of measurement points $\mathcal{S} = { \vr_1, \dots, \vr_m }$ and a set of time points $\mathcal{T} = { t_1, \dots, t_T }$.

%%%%%%%%%%%%%%%%%%%%%%%%%%%%%%%%%%%%%%%%%%%%%%%%%%%%%%%%%%%%%%%%%%%
%%         ADJOINT OPERATORS
%%%%%%%%%%%%%%%%%%%%%%%%%%%%%%%%%%%%%%%%%%%%%%%%%%%%%%%%%%%%%%%%%%%
\section{Adjoint Operators} \label{sec:Adjoint}
The adjoint operator is the generalization of the transpose of a matrix. It plays a vital role in image reconstruction techniques as it represents the backward operation, providing a path from the data domain to the image domain.

In X-ray CT, the adjoint of the Radon transform is expressed as:
\begin{align*}
\mathcal{A}_t^{H}(\vy(s,\vtheta) ; \vr) = \int_{\mathcal{S} \times \Theta} \vy(s, \vtheta) \delta(s - \langle \vr, \vn(\vtheta) \rangle ) \, \mathrm{d}s  \, \mathrm{d}\vtheta,
\end{align*}
where $\vw : \mathcal{S} \times \Theta \rightarrow \R$ is the function on the projection space, $\mathcal{S}$ denotes the set of distances $s \in \R$, and $\mathcal{A}$ denotes the set of Euler angles $\vtheta \in \R^{d-1}$. The adjoint operator integrates over the product space $\mathcal{S} \times \Theta$.

For MRI, the adjoint operator of the FTM is given by:
\begin{align*}
\mathcal{A}_t^{H}(y(\vk); \vr) = \int_{-\infty}^{\infty} y(\vk) \exp(2\pi i \left\langle \vk, \vr \right\rangle ) \mathrm{d} \vk,
\end{align*}
where $y(\vk)$ represents the data in the k-space domain, and $\vr \in \R^d$ represents the spatial coordinates in the spatial domain. The adjoint operator is essentially the inverse Fourier transform, with the complex exponential term having a positive exponent, as opposed to the negative exponent in the forward operator.

%%%%%%%%%%%%%%%%%%%%%%%%%%%%%%%%%%%%%%%%%%%%%%%%%%%%%%%%%%%%%%%%%%%
%%%         EXTENSIONS
%%%%%%%%%%%%%%%%%%%%%%%%%%%%%%%%%%%%%%%%%%%%%%%%%%%%%%%%%%%%%%%%%%%
\section{DSS Extensions}
\label{sec:Extensions}
The DSS framework can be extended to account for temporal changes in the object's density by incorporating a time-varying attenuation coefficient, $\vu$. This adjustment is reflected in the following mathematical form:
\begin{align*}
    \{ \valpha^\star, \vu^\star \} \in \underset{\valpha \, \in \, \R^k, \vu \in \R^T }{\argmin} \left\lbrace \sum_{t=1}^{T} \| \mathcal{A}_t \! \left( u_t h_\epsilon \! \left( \Psi_t \valpha \right) \right) - \vy_t \|^2 \right\rbrace
\end{align*}
where $\vu$ is a vector of attenuation coefficients. The attenuation coefficient at time $t$ is modeled as $\vx(t) = u_t h_\epsilon(\Psi_t \valpha)$.

In scenarios where the images are discrete with multiple gray-levels, the DSS framework can be extended to account for this using a discrete Heaviside function:
\begin{align*}
    \valpha^\star & \in \underset{\valpha  \in  \mathbb{R}^k}{\argmin} \left\lbrace \sum_{t=1}^{T} \| \mathcal{A}_t \left( I_t (\valpha) \right) - \vy_t \|^2 \right\rbrace, \\[1ex]
    \mbox{with} \quad I_t(\valpha) &= u_1 h_\epsilon(\Psi_t \valpha - u_1) \, \, + \\
    & \qquad \sum_{p=2}^{m} (u_p - u_{p-1}) h_\epsilon(\Psi_t \valpha - (u_p - u_{p-1})),
\end{align*}
where, $h_\epsilon(\cdot)$ is the Heaviside step function, $\valpha$ is the unknown variable, $\Psi_t$ is the operator at time $t$, and $u_1, \ldots, u_m$ are the known material attenuation coefficients in increasing order. This formulation effectively allows the framework to handle images with multiple gray-levels representing different material intensities.

Additionally, to tackle extreme noise scenarios, we need to impose a geometric constraint on the reconstructed shape. To achieve this, regularization can be applied to the boundaries by integrating the Dirac-delta function. This integration introduces a penalty term that aims to minimize the circumference of the object, as shown in the following equation:
\begin{align*}
    \valpha^\star \in \underset{\valpha \, \in \, \R^k}{\argmin} \left\lbrace \sum_{t=1}^{T}  \| \mathcal{A}_t \! \left( h_\epsilon(\mPsi_t \valpha) \right) - \vy_t \|^2 + \lambda \| \delta_\epsilon \! \left( \Psi_t \valpha \right) \|^2 \right\rbrace,
\end{align*}
where $\delta_\epsilon$ denotes the Dirac-delta function, which is specifically used here to estimate the circumference of the shape, thereby serving as a regularization mechanism to penalize larger circumferences. The coefficient $\lambda$ is a regularization parameter controlling the balance between data fidelity and this circumference-based regularization \cite{mumford1989optimal, chan2001active}. By carefully selecting the regularization parameter, the model can effectively handle extreme noise scenarios, leading to more accurate and stable shape reconstructions.

Hence, the proposed compressed dynamic shape sensing framework is a powerful tool for reconstructing the shape of dynamic objects from a time-series of measurements. It can be extended to account for changes in density and discrete gray-levels, and can incorporate regularization in the presence of extreme noise scenarios. However, it still requires a considerable amount of computational resources to find an optimal solution.

\section*{Acknowledgment}
This work was supported by the Dutch Research Council (NWO, project number 613.009.106 and 639.073.506). Ajinkya Kadu also acknowledges financial support from ERC Consolidator Grant Number 815128 REALNANO. The authors thank Math\'e Zeegers, Nick Luiken, Tristan van Leeuwen, Sara Bals, and Matteo Ravasi for useful discussion and comments.

\ifCLASSOPTIONcaptionsoff
  \newpage
\fi


\end{document}